\documentclass[a4paper]{llncs}

\usepackage{color}
\usepackage{amssymb} 
\usepackage{graphicx}
\usepackage[ngerman,english]{babel}
\usepackage[lined,algonl,boxed]{algorithm2e}
\usepackage[latin1]{inputenc}

\newcommand{\PP}{P}
\newcommand{\C}{C}
\newcommand{\D}{{\mathcal D}}

\newcommand{\X}{X}

\newcommand{\alldiff}{\texttt{alldiff}}

\newcommand{\CGMC}{CG^{MC}}

\begin{document}

\pagestyle{plain}

\title{Decomposition Techniques for Subgraph Matching}

\author{Stéphane Zampelli$^1$, Martin Mann$^2$, Yves Deville$^1$,
  Rolf Backofen$^2$} 

\institute{
 $^{(1)}$ Université catholique de Louvain, \\
Department of Computing Science and Engineering, \\ 
2, Place Sainte-Barbe 1348 Louvain-la-Neuve (Belgium) \\
\email{\{stephane.zampelli,yves.deville\}@uclouvain.be}  \\
$^{(2)}$ Albert-Ludwigs-University Freiburg, \\
Bioinformatics Group, Institute of Computer Science, \\
Georges-Koehler-Allee 106, D-79110 Freiburg (Germany)\\
\email{\{mmann,backofen\}@informatik.uni-freiburg.de}
}

\maketitle

\begin{abstract}

In the constraint programming framework, state-of-the-art static and dynamic 
decomposition techniques are hard to apply to problems with complete initial 
constraint graphs. For such problems, we propose a hybrid approach of these 
techniques in the presence of global constraints. In particular, we solve the 
subgraph isomorphism problem. Further we design specific heuristics for this 
hard problem, exploiting its special structure to achieve decomposition. The 
underlying idea is to precompute a static heuristic on a subset of its 
constraint network, to follow this static ordering until a first problem 
decomposition is available, and to switch afterwards to a fully propagated, 
dynamically decomposing search. Experimental results show that, for sparse 
graphs, our decomposition method solves more instances than dedicated, 
state-of-the-art matching algorithms or standard constraint programming 
approaches.

\end{abstract}

\section{Introduction}

Graph pattern matching is a central application in many
fields~\cite{Conte2004} and can be successfully modeled using
constraint programming~\cite{LV02,Rud98,approcp2005}. Here, we stress
how to apply decomposition techniques to solve the Subgraph
Isomorphism Problem (SIP) in order to outperform the dedicated
state-of-the-art algorithm.

Decomposition techniques are an instantiation of the divide and
conquer pa\-ra\-digm to overcome redundant work for independent
partial problems. A constraint problem (CSP) can be associated with
its constraint network, which represents the active constraints
together with their relationship. During search, the constraint
network looses structure as variables are instantiated and
constraints entailed by domain propagation. The constraint network
can possibly consist of two or more independent components,
leading to redundant work due to the repeated computation and
combination of the corresponding independent partial solutions. The
key to solve this is \textit{decomposition} that consists of two
steps.  The first step detects the possible problem decompositions,
by examining the underlying constraint network for independent
components. The second step exploits these independent components by
solving the corresponding partial CSPs independently, and combines
their solutions without redundant work. Decomposition can occur at
any node of the search tree, i.e. at the root node or dynamically
during search. In constraint programming, decomposition techniques
have been studied through the concept of AND/OR search
\cite{mateescu2007}. AND/OR search is sensitive to problem
decomposition, introducing search subtree combining AND nodes as an
extension to classical OR search nodes. The size of the minimal
AND/OR search tree is exponential in the tree width while the size of
the minimal OR search tree is exponential in the path width, and is
never worse than the size of the OR tree search.

The check for decomposition is usually done in one of two ways.
Either, only the initial constraint network is statically analysized,
resulting in a so called pseudo-tree. This structure encodes both,
the static search heuristic and the information when a subproblem is
decomposable~\cite{Dechter:Mateescu:impact:CP2004}. Another
possibility is to consider the dynamic changes of the constraint
network by analyzing it at each node during the
search~\cite{Dechter-Mateescu-aij07-dynamicANDOR}. Such a dynamic
approach is better suited if a strong constraint propagation (e.g. by
AC) is present but obviously to the cost of additional computations.

A major problem of decomposition techniques are their problem
specificity.  Without good heuristics, decomposition may occure
seldom or very late such that the computational overhead for checking
etc. is too high for an efficient application. Nevertheless, some
approaches have been shown to be more general by applying dedicated
algorithms, e.g. graph separators or cycle cutset
conditioning~\cite{KARYPIS-1998,mateescu2007,Mateescu:Dechter:ANDORcutset:IJCAI2005}.

However, those (usually static) algorithms fail to compute good
heuristics on problems with global constraints, which have an
initially complete constraint graph. Indeed, such algorithms
presuppose a sparse constraint graph. In the subgraph isomorphism
problem (SIP), for example, the initial constraint graph is complete
due to the presence of a global \alldiff-constraint. This prevents
cycle cutset and graph separator algorithms to be applied. A further
drawback of a static analysis is the non-predictable decomposability
of the constraint network achieved by constraint entailment through
propagation. To exploit this, a dynamic analysis of the problem
structure during the search is necessary.  This is of high importance
for SAT-~\cite{Li:Beek:SAT-decomp:2004} and
CSP-solving~\cite{Dechter-Mateescu-aij07-dynamicANDOR}. Unfortunately,
a dynamic analysis requires significant additional work that slows
down the search process once more.

In this paper we show how to overcome those shortcomings by combining
static and dynamic decomposition approaches to take advantage of
decomposition for the hard problem of SIP. A combination yields a
balance between the fast static analysis and the needed full
propagation exploited by dynamic search strategies in the presence of
global constraints. The underlying idea is to follow the static
ordering until a first problem decomposition is available (or likely)
and to switch afterwards to a full propagated decomposing search. For
the later, we consider only a binary constraint representation inside
the constraint network in order to compute a good
decomposition-enforcing heuristic.  As shown in the experiments, this
idea is a key point for an efficient application of a decomposing
search (as AND/OR) for the SIP.

To face the problem of graph pattern matching~\cite{Conte2004} many
different types of algorithms have been proposed, ranging from
general methods to specific algorithms for particular types of
graphs. The state-of-the-art approach is the dedicated
\emph{VF}-algorithm, freely available in the C++ \texttt{vflib}
library \cite{Cordella2001}. In constraint programming, several
authors \cite{LV02,Rud98} have shown that graph matching can be
formulated as a CSP problem, and argued that constraint programming
could be a powerful tool to handle its combinatorial complexity.  Our
modeling \cite{approcp2005} is based on these works.  In
\cite{approcp2005}, we showed that a CSP approach is competitive with
dedicated algorithms over a graph database representing graphs with
various topologies.  Regarding decomposition, Valiente and al.
\cite{Val1997} have shown how to use decomposition techniques in
order to speed up subgraph homeomorphism. \cite{Val1997} states that,
if the initial pattern graph is made of several disconnected
components, then matching each component separately is equivalent to
matching all of them together. Specific algorithms are also
demonstrated.  Our work can be seen as an extension to this work. We
consider the subgraph isomorphism problem instead of the subgraph
homeomorphism problem. The latter case is easier as the constraint
graph is made only of the initial pattern graph. Moreover, we apply
the decomposition dynamically when \cite{Val1997} decomposes only
statically on the initial pattern graph.

\vspace{-1em}
\subsubsection{Objectives and results - }
In this paper we study the limits of the direct application of
state-of-the-art (static and dynamic) decomposition techniques for
problems with global constraints; we show that such a direct
application is useless for SIP. We develop a hybrid decomposition
approach for such problems and design specific search heuristics for
SIP, exploiting the structure of the problem to achieve
decomposition. We show that the CP approach using the proposed
decomposition techniques outperforms the state-of-the-art algorithms,
and solves more instances on some classes of problems (sparse
instances with many solutions).

The paper is structured as follows.  Section \ref{sec:defs}
introduces a decomposition method able to detect decomposition at any
stage during the search.  In Section \ref{sec:SIP}, the proposed
decomposition method is applied and specialized to SIP.  Experimental
results assessing the efficiency of our approach are presented in
Section \ref{sec:results}.  Section \ref{sec:conc} concludes the
paper.

\section{Decomposition}
\label{sec:defs}

In this section we show how to define and detect decomposition during
search.  Sections \ref{sec:prelim} and \ref{sec:decomp} define a
decomposition method able to detect decomposition at any state during
search, considering that we do not know a priori when decomposition
occurs.  Section \ref{sec:AND/OR} shows that our method is able to compute the
same decompositions than the AND/OR search framework
\cite{mateescu2007}, where the search is precomputed on a graph
representation of the constraint network, and decomposition events
are known in advance.  The AND/OR search method has shown to be very
attractive for a large classes of constraint networks. But as we will
see in Section \ref{sec:SIP}, our method is suited for the SIP
while the AND/OR method is not applicable because the decomposition
events cannot be precomputed.

\subsection{Preliminary}
\label{sec:prelim}

A \emph{Constraint Satisfaction Problem (CSP)} $P$ is a triple
$(\X,\D,\C)$ where $\X=\{x_1,\dots,x_n\}$ is a set of variables,
$\D=\{D_1,\dots,D_n\}$ is a set of domains (i.e. a finite set of
values), each variable $x_i$ is associated with a domain $D_i$, and
$\C$ is a finite set of constraints with $scope(c) \subseteq \X$ for
all $c \in \C$, where $scope(c)$ is the set of variables involved in
the constraint $c$. A constraint $c$ over a set of variables defines
a relation between the variables. A \emph{solution} of the CSP is an
assignment of each variable in $\X$ to one value in its associated
domain so that no constraint $c \in \C$ is violated. We denote
$Sol(P)$ the set of solutions of a CSP $P$.

A \emph{partial CSP} $\hat P$ of a CSP $P \equiv (X,\D,C)$ is a CSP
$(\hat X, \hat \D, \hat C)$ where $\hat X \subseteq X$, $\forall \hat
D_k \in \hat \D: \hat D_k \subseteq D_k$ and $\hat\C \subseteq\C$.
Note that since $\hat P$ is a CSP, we have $scope(\hat{c}) \subseteq
\hat{\X}$ for all $\hat{c} \in \hat{\C}$.

\subsection{Decomposing CSPs and graphs}
\label{sec:decomp}

This subsection defines the notion of decomposition for a CSP.
A CSP is decomposable into partial CSPs if the CSP and its decomposition 
have the same solutions.

\begin{definition} 
A CSP $P$ is \emph{decomposable} in partial CSPs $P_1,\dots,P_n$ iff : 
\begin{itemize}
\item $\forall\; s \in Sol(P): \exists\; s_1,\dots,s_k \in Sol(P_1),\dots,Sol(P_k) : s=\cup_{i \in [1,k]} s_i$ 
\item $\forall\; s_1,\dots,s_k \in Sol(P_1),\dots,Sol(P_k): \exists\; s \in Sol(P) : s=\cup_{i \in [1,k]} s_i$.
\end{itemize}
\end{definition}

This general definition of decomposition can be instantiated to two
practical cases. The first definition corresponds to the direct
intuition of a decomposition: a CSP is decomposable if it can be
split into disjoint partial CSPs.
It is called 0-decomposability as no variable are shared between the
partial CSPs. 

\begin{definition}
\label{def:0-decomp}
A CSP $P=(X,\D,C)$ is \emph{0-decomposable} in partial CSPs
$P_1,\dots,P_n$ with $P_i=(X_i,\D_i,C_i)$ iff $\forall\; 1 \leq i<j \leq n : 
X_i \cap X_j = \emptyset$, $\cup_{i \in [1,k]} X_i
= X$, $\cup_{i \in [1,k]} \D_i = \D$, $\cup_{i \in [1,k]} C_i = C$.
\end{definition}

The second definition finds more decompositions by allowing the
partial CSPs to have instantiated variables in common.
It is called 1-decomposability as variables shared between the
partial CSPs have a domain of size 1.

\begin{definition} 
\label{def:1-decomp}
A CSP $P=(X,\D,C)$ is \emph{1-decomposable} in partial CSPs
$P_1,\dots,P_k$ with $P_i=(X_i,\D_i,C_i)$ iff $\forall\; 1 \leq i<j \leq n :
x \in (X_i \cap X_j) \Rightarrow |D_x|=1$, $\cup_{i
\in [1,k]} X_i = X$, $\cup_{i \in [1,k]} \D_i = \D$, $\cup_{i \in
[1,k]} C_i = C$.
\end{definition}

The relationship with the general definition is direct. If a CSP $P$
is \emph{0-decomposable} or \emph{1-decomposable} in partial CSPs
$P_1,\dots,P_k$, then $P$ is decomposable in partial CSPs
$P_1,\dots,P_k$.  From Definitions \ref{def:0-decomp} and
\ref{def:1-decomp}, it follows further~:

\begin{property}
If a CSP $P=(\X,\D,\C)$ is 0-decomposable in $\PP_1,\dots,\PP_k$, then $P$ is 
1-decomposable in $\PP_1,\dots,\PP_k$. Further $P$ might be 1-decomposable in
$\PP'_1,\dots,\PP'_{k'}$ with $k' \geq k$ via overlapping partial
problems~$\PP'_i$. 
\end{property}

Redundant computation during CSP-solving is performed whenever a CSP is 0-~or 
1-decomposable into $k$~partial CSPs $\PP_1,\dots,\PP_k$. For instance, 
if the solutions of $\PP_1$ are computed first, then for each solution of 
$\PP_1$ repeatedly all solutions of $\PP_2,\dots,\PP_k$ are computed. 
Therefore, $\PP_2,\dots,\PP_k$ are solved $|Sol(\PP_1)|$ times and 
this overhead can be exponential in the size of the CSP. This can be avoided by 
solving the partial problems independently. The necessary detection of the 
CSP-decomposition into independent partial CSPs can be performed through the 
concept of constraint graphs.

A \textit{graph} $G=(V,E)$ consists of a \textit{vertex/node set} $V$ and an 
\textit{edge set} $E \subseteq V \times V$, where an edge $(u,v)$ is a pair of 
nodes. The vertices $u$ and $v$ are the endpoints of the edge $(u,v)$. We 
consider directed and undirected graphs. A \textit{subgraph} of a graph 
$G=(V,E)$ is a graph $G'=(V',E')$ with $V'\subseteq V$ and $E'\subseteq E$ such 
that $\forall_{(u,v) \in E'} : u,v \in V'$. A graph $G$ is said to be 
\textit{singly connected} if and only if there is at most one simple path 
between any two nodes in $G$.

\begin{definition}
\label{def:CG}
\begin{sloppy}
The \emph{constraint graph} of a (partial) CSP $P = (\X,\D,\C)$ is an
undirected graph $G^P = (V,E)$ where $V = \X$ and $E=\{(x_i,x_j)
\;|\; \exists\; c \in \C : x_i,x_j \in scope(c) \}$.
\end{sloppy}
\end{definition}

Note that all variables in the scope of one constraint form a clique in
$G^P$. This constraint graph is also called the \textit{primal graph}
\cite{dechter2003}. There is a standard syntactic way of decomposing
a CSP, based on its constraint graph.

\begin{definition} \label{def:graph-dec}
A graph $G=(V,E)$ is \emph{decomposable} into $k$ subgraphs $G_1,\dots,G_k$ iff 
$\forall_{1\leq i<j\leq k} : V_i \cap V_j = \emptyset$, $\cup_{i \in [1,k]} V_i 
= V$, and $\cup_{i \in [1,k]} E_i = E$.
\end{definition}

Property \ref{pro:decomp1} shows that one has to compute disjoint
components of the constraint graph to detect independent CSPs.  This
can be done in linear time.

\begin{property}
\label{pro:decomp1}
Given a CSP $P=(\X,\D,\C)$ with its constraint graph $G$,
for all $k \geq 1$,
the constraint graph $G$ of $P$ is decomposable in $G_1,\dots,G_k$,
iff $P$ is 0-decomposable in $P_1,\dots,P_k$
iff $P$ is 1-decomposable in $P'_1, \dots, P'_m$ with $m\geq k$.
\end{property}
\textbf{Proof -} The first iff is straightforward.  For the second
iff, we can construct a 1-decomposition $P_1,\dots,P_m$ of $P$ from a
decomposition $G_1,\dots,G_k$ of $G$, with $m \geq k$.  The
construction is described for the case $k=1$ (i.e. $P_1=P$), and can
be easily generalized.  Let $G=(V,E)$ be the graph constraint of $P$.
Let $V_s=\{x \in V \;|\; |D_x|=1 \}$.  Transform $G$ into $G'$ where
$G'$ is the graph $G$ without variables with a singleton domain. More
formally, $G'=(V',E')$ with $V'= V \setminus V_s$ and $E'=(V' \times
V') \cap E$.  Suppose $G'$ is decomposable into $G'_1, \dots,G'_m$
$(m\geq 1)$.  Then, nodes associated to variables with a singleton
domain and their associated edges are added to the $G'_i$, giving
$G^{1}_i=(V^{1}_i,E^{1}_i)$.  More formally $G^1_{i}=(V^{1}_{i},
E^{1}_{i})$ where $V^{1}_{i}= V'_{i} \cup V_s$ and $E_i^{1} =
(V^{1}_i \times V^{1}_i) \cap E$.  The graphs $G^1_{1},\dots,G^1_{m}$
are the constraint graphs of the partial CSPs $P_i$ of the
1-decomposition of $P$. $\blacksquare$

The above property is especially useful when $k=1$. 
In this case, the 0-decomposition does not decompose the CSP,
while 1-decomposition may decompose it.

\subsection{Relationship with AND/OR search tree}
\label{sec:AND/OR}

Another approach to define decomposable CSPs is to use the concept of
AND/OR search spaces defined with pseudo-trees \cite{mateescu2007}.

\begin{definition}
Given an undirected graph $G=(V,E)$, a directed rooted tree $T=(V,E')$
defined on all its nodes is called \emph{pseudo-tree} of $G$ if any arc of
$E$ which is not included in $E'$ is a back-arc, namely it connects a
node to an ancestor in $T$.
\end{definition}

\begin{definition}
Given a CSP $P=(\X,\D,\C)$, its constraint graph $G^P$ and a pseudo-tree
$T^P$ of $G^P$, the associated AND/OR search tree has alternating levels
of OR nodes and AND nodes. The OR nodes are labeled $x_i$ and
correspond to variables. The AND nodes are labeled $<x_i,v_k>$ and
correspond to assignment of the values $v_k$ in the domains of the
variables. The root of the AND/OR search tree is an OR node, labeled
with the root of the pseudo-tree $T^P$.  The children of an OR node
$x_i$ are AND nodes labeled with assignments $<x_i,v_k>$, consistent
along the path from the root.  The children of an AND node
$<x_i,v_k>$ are OR nodes labeled with the children of variable $x_i$
in $T^P$.
\end{definition}

Semantically, the OR states represent alternative solutions, whereas
the AND nodes represent the problem decompositions into independent
partial problems, all of which need to be solved. When the pseudo-tree is
a chain, the AND/OR search tree coincides with the regular OR search
tree.

Following the ordering induced by the given a pseudo-tree $T^P$ of the 
constraint graph of a CSP~$P$, the notion of 1-decomposability coincides with 
the decompositions induced by an AND/OR search.

\begin{property}
Given a CSP $P=(\X,\D,\C)$, a pseudo tree $T^P$ over the constraint graph
of $P$ and a path $p$ of length $l$ ($l \geq 1$) from the root node
of $T^P$ to an AND node~$p_l$, the CSP $P$ where all variables in the path $p$ are
assigned is 1-decomposable into $\PP_1,\dots,\PP_k$ where $k$ is the number of
OR~successors in $T^P$ of the end node $p_l$.

\textbf{Proof -} 
Let $y_1,\dots,y_k$ ($k \geq 1$) be the OR successor
nodes of the end node $p_l$ in $T^P$.  We note $tree(y_i)$ the tree
rooted at $y_i$ in $T^P$.  Let $X_s=\{v \in \X | v \in p \}$.  Then
build the partial CSPs $\PP_i=(\X_i,\D_i,\C_i)$ $(i \in [1,k])$:
\begin{eqnarray*}
\X_i & = & X_s \cup \{ v \in \X \;|\; v \in tree(y_i) \} \\ 
\D_i & = & \{ D_x \in \D \;|\; x \in \X_i \} \\
\C_i & = & \{ c \in \C \;|\; scope(c) \subseteq \X_i \}. \\
\end{eqnarray*}
It is clear that $\cup_{i \in [1,k]} \C_i=\C$ since there exists no
constraint between two different $tree(y_i)$ in $T^P$, by definition of
a pseudo tree. $\blacksquare$
\end{property}

As will be explained in the next section, neither static nor dynamic
AND/OR search is suited for our particular problem.  In SIP, the
constraint graph is complete, and thus the pseudo tree is a chain,
leading to an AND/OR search tree equivalent to an OR search tree.
However the CSP $P$ becomes 1-decomposable during search and a
\emph{dynamic} framework is needed in order to check decomposition on
any state during the search. But this is computationally very
expensive as we will show in Section~\ref{sec:results}.

\section{Applying decomposition to SIP}
\label{sec:SIP}

\subsection{Subgraph Isomorphism Problem Definition}
\label{sec:SIPD}

A \emph{subgraph isomorphism problem} between a pattern graph
$G_p=(V_p,E_p)$ and a target graph $G_t=(V_t,E_t)$ consists in
deciding whether $G_p$ is isomorphic to some subgraph of $G_t$. More
precisely, one should find an injective function $f: V_p\rightarrow
V_t$ such that $\forall (u,v) \in E_p :\; (f(u),f(v)) \in
E_t$. This NP-Hard problem is also called subgraph monomorphism
problem or subgraph matching in the literature. The function $f$ is
called a {\em subgraph matching function}.  We assume the graphs are
directed.  Undirected graphs are a particular case where undirected
arcs are replaced by two directed arcs.

The CSP model $P=(\X,\D,\C)$ of subgraph isomorphism should represent
a total function $f : V_p \rightarrow V_t$. This total function can
be modeled with $\X={x_1, \dots, x_n}$ with $x_i$ a FD variable
corresponding to the i$^{th}$ node of $G_p$ and $D(x_i)=V_t$. The
injective condition is modeled with an \alldiff$(x_1,\dots,x_n)$
global constraint. The isomorphism condition is translated into a set of
$n$~k-ary constraints $MC_i \equiv (x_i,x_j) \in E_t$ for all $x_i
\in V_p$. Given
the above modelling, the constraint graph of the CSP, called the SIP
constraint graph, is the graph $G^P=(V^P,E^P)$ where $V^P=\X$ and
$E^P=E_p \cup E_{\neq}$. Note, $E_p$ is representing all propagations
of the $MC_i$ constraints while $E_{\neq}$ depicts the global
\alldiff-constraints, i.e. a clique ($E_{\neq}=V_p\times
V_p$). 
Therefore, the SIP-CSP consists of global constraints only
that would prevent decomposition using a static AND/OR search. 
Implementation, comparison with dedicated algorithms, and
extension to subgraph isomorphism and to graph and function
computation domains can be found in \cite{approcp2005}.

\subsection{Decomposing SIP}
\label{sec:SIPCG}

This subsection explains how to decompose the SIP problem.
We first show why static AND/OR search fails by studying the SIP constraint
graph.

\paragraph{\bfseries Static AND/OR Search:}
Because of the \alldiff-constraint, the SIP constraint graph
corresponds to the complete graph $K_{|V_p|}$.  The pseudo-tree
computed on the constraint graph of any SIP instance is a chain,
detecting no decomposition at all. Moreover, the initial SIP
constraint graph is not 1-decomposable. Therefore a static analysis
of the SIP-CSP yields no decomposition at all and is not applicable.

Decomposition seems difficult to achieve. However, as variables are
assigned during search, 1-decomposition may occur at some nodes of
the search tree.  A dynamic detection of 1-decomposition at different
nodes of the search tree gives a first way of detecting decomposition
for the SIP.

\paragraph{\bfseries Dynamic AND/OR Search:}
A dynamic analysis of the SIP constraint graph, as done for dynamic
AND/OR search, takes care of possible constraint entailments and
propagation results.  It is therefore very usefull for a strongly
propagated CSP. The main drawback is the slow down due to the
additional propagation and dynamic decomposition checks. Further, the
SIP constraint graph is still a complete one and does not allow for
decomposition.

Our 1-decomposition removes assigned variables in the decomposition
process.  One could also remove entailed constraints, leading thus to
more decomposition.  This can easily be done for the
\alldiff-constraint by removing an edge $(x_i,x_j)\in E^P$
representing $x_i \neq x_j$ when $D_i \cap D_j=\emptyset$ $(i\neq
j)$.  In the following, we redefine the constraint graph of a SIP as
a constraint graph for the morphism constraints together with a
dynamic constraint graph of the \alldiff-constraint.

\begin{definition} \label{def:sip-graph}
Given the CSP $P=(\X,\D,\C)$ of a SIP instance, its \emph{SIP constraint
graph} is the undirected graph $G=(V,E^{MC} \cup E^{\neq} )$, where $V=\X$,
$E^{MC}=\{(x_i,x_j)\in E_{p}\;|\;x_i,x_j\in\X\}$ and $E^{\neq}=\{(i,j) \in \X \times \X \;|\; D_i
\cap D_j = \emptyset \}$.
\end{definition}

Given the particular structure of a SIP constraint graph, it is possible to 
specialize and simplify the detection of 1-decomposition.

\begin{property}
\label{pro:pgbreaks}
Let $P=(\X,\D,\C)$ be a CSP model of a SIP instance, and let
$G=(V,E^{MC} \cup E^{\neq})$ be its SIP constraint graph.  Let
$M=(V',E')$ be the constraint graph without assigned variables,
i.e. with $V'=\{ x \in \X \;|\; |D_x|>1 \}$ and $E'=(V' \times V') \cap
E^{MC}$.  
Then $P$ is 1-decomposable into $P_1,\dots,P_m$ iff $M$ is
decomposable into $M_1,\dots,M_m$ and $D(M_i) \cap D(M_j)=\emptyset$
$(1 \leq i < j \leq m)$ with $D(M_i)$ the union of the domains of the
variables associated to the nodes of $M_i$.
\end{property}

The above property states that the decomposition of
$M$ is a necessary condition.
We can therefore design heuristics leading to the decomposition
of $M$, hence in some cases in the decomposition of $P$.

A direct approach consists in detecting 1-decomposition at each node
of the search tree.  When the CSP becomes 1-decomposable in partial
CSPs, those are computed separately in AND nodes.  As show in the
experimental section, this strategy proves to be much slower than a
standard OR search tree.  The reason is twofold:
\begin{enumerate}
\item Decomposition is tested at every node of the search tree.
Starting from the root node is useless, as a lot of computation time is lost.
\item There is no guarantee that a decomposition will occur.
\end{enumerate}

Based on this observation, we present a \emph{hybrid approach}
combining the best of the static and dynamic strategy.

\paragraph{\bfseries The Hybrid Approach:} 
As stated before, even a dedicated dynamic AND/OR search, checking
for decomposition on the reduced constraint graph only, is not fast
enough to compete with state-of-the-art SIP-solvers as implemented in
the \texttt{vflib} library. Therefore, we suggest a hybrid approach
in order to fix this. The idea is as follows:
\begin{enumerate}
  \item calculate a static pseudotree heuristic on the reduced constraint graph
  \item apply a forward checking search following the pseudotree up
  to the first branching or until a fixed number of variables is
  assigned
  \item switch the strategy to dynamic AND/OR search with full AC-propagation
\end{enumerate}

This ensures, that the expensive dynamic approach is first used when
a decomposition is available or at least likely after full
propagation. Up to that moment, a cheap forward checking approach is
used for a fast inconsistency check and a strong reduction of the
reduced constraint graph.

In the following, we will give two dedicated heuristics we have
applied in the preliminary forward checking procedure.

\subsection{Heuristics}
\label{sec:heuristics}

We now present two heuristics based on Property \ref{pro:pgbreaks}
aiming at reducing the number of decomposition tests, and favoring
decomposition.  The general idea is to first detect a subset of
variables disconnecting the morphism constraint graph into disjoint
components as it is a necessary condition for 1-decomposability.  The
search process will first distribute over these variables.  The test
of 1-decomposition is performed when all these variables are
instantiated. It is also performed at the subsequent nodes of the
search tree.

\subsubsection{The cycle heuristic (h1)}

The objective of the cycle heuristic is to find a set of nodes $S$ in
the morphism graph $\CGMC=(X,E^{MC})$ (see Def.~\ref{def:sip-graph}) such that the graph without those
nodes is simply connected.  When the variables associated to $S$ are
assigned, any subsequent assignment will decompose the morphism
graph. Finding the minimal set of nodes is known as the minimal cycle
cutset problem and is a NP-Hard problem \cite{cutset}.  We propose
here a simple linear approximation that returns the nodes of the
cycles of the graph.  Algorithm \ref{algo:heuristics} runs in
$O(|V_p|)$.  The effectiveness of such a procedure on different
classes of problems is shown in the
experimental section. One of the main advantage is its simplicity.

\begin{algorithm}
  \SetKwData{G}{G}
  \SetKwFunction{OutDegree}{OutDegree}
  \SetKwInOut{Input}{input}
  \SetKwInOut{Output}{output}
  \caption{Selection of the body variables.} 
  \Input{$G=(X,E)$ the $\CGMC$}
  \Output{The nodes of the cycles of $G$}
  \BlankLine
   $All \leftarrow X$ \\
   $T \leftarrow \emptyset$ \\
\While{($\exists\; n \in X \;|\; Degree(n)==1$) }
{
	$T \leftarrow T \cup \{n\}$ \\
	remove node $n$ from $G$ \\
}
return $All \setminus T$
  \label{algo:heuristics}
\end{algorithm}

\subsubsection{Using graph partitioning (h2)}

Graph partitioning is a well-known technique that allows hard graph
problems to be handled by a divide and conquer approach.  In our
context, it can be used to separate the morphism constraint graph
into two graphs of equal size.

\begin{definition}
Given a graph $G=(V,E)$, a $k$-graph partitioning of $G$ is a
partition of V in $k$ subsets, $V_1,\dots,V_k$, such that $V_i \cap
V_j=\emptyset$ for $i \neq j$, $\cup_{i} V_i=V$, and the number of
edges of $E$ whose incident vertices belong to different subsets is
minimized (called the \emph{edgecut}).
\end{definition}

Based on the edgecut of the morphism constraint graph, we can easily
deduce a subset variables.

\begin{definition}
Given a 2-graph partitioning of $G$, a \emph{nodecut} is a set of
nodes containing one node of each edge in the cutset.
\end{definition}

Finding a minimum edgecut is a NP-Hard problem for $k \geq 3$, but
can be solved in polynomial time for $k=2$ by matching (see
\cite{np-problems:garey_johnson:1990}, page 209).  However we use a
fast local search approximation \cite{metis}, as the exact minimum
subset is not needed.

\section{Experiments}
\label{sec:results}

\subsubsection{Goals -} 
The objective of our experiments is to compare our decomposition method
on different classes of SIP with standard CSP models as well as \texttt{vflib},
the standard and reference algorithm for subgraph
isomorphism \cite{Cordella2001}.  We also compare our decomposition method with
standard direct decomposition.  The different
heuristics presented in Section 3.5 are also tested.

\subsubsection{Instances -} 
The instances are taken from the \texttt{vflib} graph database
described in \cite{graphdb}.
There are several classes of randomly generated graph, random graphs,
bounded graphs and meshes graphs.  The target graphs has a size $n$
and the relative size of the pattern is noted $\alpha$.  For random
graphs, the target graph has a fixed number of nodes $n$ and there is
a directed arc between two nodes with a probability $\eta$.  The
pattern graph is also generated with the same probability $\eta$, but
its number of nodes is $\alpha n$.  If the generated graph is not
connected, further edges are added until the graph is connected.  For
random graphs, $n$ takes a value in
$[20,40,80,100,200,400,800,1000]$, $\eta$ in $[0.01,0.05,0.1]$, and
$\alpha$ in $[20\%,40\%,60\%]$. There are thus 69 classes of randomly
connected graphs.   In a class of instances denoted as
\texttt{si2-r001-m200}, we have $\alpha=20\%$,
$\eta=0.01$, and $n=200$ nodes.

Mesh-$k$-connected graphs are graphs where each node is connected
with its $k$ neighborhood nodes.  Irregular mesh-$k$-connected graphs
are made of a regular mesh with the addition of random edges
uniformly distributed. The number of added branches is $\rho n$.  For
random graphs, $n$ can take a value in $[16, \dots, 1096]$, $k$ in
$[2,3,4]$, and $\rho$ in $[0.2,0.4,0.6]$. In an irregular mesh-connected
class of instances denoted as
\texttt{si2-m4Dr6-m625}, we have $\alpha=20\%$, $k=4$, $\rho=0.6$
and $n=625$ nodes.

One hundred graphs are generated for each class of instances.  For
random graphs, we also generated 100 additional instances where the
target graph has 1600 nodes, for each possible value of $\eta$ and
$\alpha$. We used the generator freely available from the graph
database, following the methodology described in \cite{graphdb}.

\subsubsection{Models  -} 
Several models were considered for the experiments. First of all, we use the 
available implementation of \texttt{vflib}.  Then classical CP models are used, 
called \texttt{CPFC} and \texttt{CPAC}. The model \texttt{CPFC} is a model 
where all the constraints use forward checking and the variable selection 
selects the first variable which is involved in the maximum number of 
constraints (called \texttt{maxcstr}) using minimal domain size as tiebreaker. 
The model \texttt{CPAC} is similar except it uses an arc consistent version of 
the \texttt{MC} constraint.

The model \texttt{CP+Dec} waits for 30\% of the variables to be instantiated
following a variable selection policy, called {\tt minsize}), selecting the
(uninstanciated) variable with the smallest domain.  It then tests at each node
of the search tree if decomposition occurs using a \texttt{maxcstr} variable
selection.  The model \texttt{CP+Dec+h1} uses the cycle heuristics; once the
nodes belonging to the cycles of the pattern graph are instantiated using a
{\tt minsize} variable selection policy (up to 30\% of the size of the pattern), decomposition is tested at each node
of the search tree and follows a \texttt{maxcstr} variable selection.  The
model \texttt{CP+Dec+h2} uses the graph partitioning heuristics; once the
variables belonging to the nodecut set are instantiated (up to 30\% of the size of the pattern), decomposition is
tested at each node of the search tree and follows a \texttt{maxcstr} variable
selection.

\subsubsection{Setup -} 
All experiments were performed on a cluster of 16 machines (AMD Opteron(tm) 875 
2.2Ghz with 2Gb of RAM) using the implementation 
of~\cite{Mann:Tack:Will:DDSintegration:CoRR2007}. All runs are limited to a 
time bound of 10 minutes.  In each experiment, we search for all solutions.  
Experiments searching for one solution have also been done but are not reported 
here for lack of space.  These experiments lead to the same conclusions.

\subsubsection{Description of the tables  -} 
Table \ref{random} shows the results for random graphs and Table
\ref{meshes} for irregular mesh-connected graphs.  Each line
describes the execution of 100 instances from a particular class.
The column $N$ indicates the mean number of solutions among the solved
instances.  The column $\%$ indicates the number of instances
that were solved within the time bound of 10 minutes.  The column
$\mu$ indicates the mean time over the solved instances and the column
$\sigma$ indicates the corresponding standard deviation. 
The column $D$ indicates the number of instances that used
decomposition among the solved instances.  The column $\#D$ indicates
the mean number of decomposition that occurred over all solved
instances.  The column $S$ indicates the mean size of the initial
variable set computed by the heuristics \texttt{h1} or \texttt{h2}.
Table \ref{degree} gives the mean degree and its variance for the
different instances classes.  For each class of instances in Tables
\ref{random} and \ref{meshes}, the results of the best algorithms are
in bold.

\begin{table}[tb]
\caption{Randomly connected graphs, searching for all solutions.}
\begin{center}
\begin{tabular}{|l|c|ccc|ccc|ccc|}
\hline
Bench &  & \multicolumn{ 3}{c|}{vflib} & \multicolumn{ 3}{c|}{CPAC} & \multicolumn{ 3}{c|}{CPFC} \\ \hline
 & N & \% & $\mu$ & $\sigma$ & \% & $\mu$ & $\sigma$ & \% & $\mu$ & $\sigma$ \\ \hline
si2-r001-m200 & 61E+6 & 72 & 74 & 115 & 83 & 56 & 109 & 85 & 41 & 76 \\ \hline
si2-r001-m400 & 17E+8 & 2 & 248 & 118 & 10 & 106 & 156 & 7 & 288 & 177 \\ \hline
si2-r001-m800 & 28E+7 & 0 & - & - & 11 & 220 & 136 & 1 & 153 & - \\ \hline
si2-r001-m1600 & 2500 & 16 & 203 & 202 & \textbf{30} & \textbf{227} & \textbf{146} & 0 & - & - \\ \hline \hline
si6-r01-m200 & 1 & 100 & 2 & 3 & 100 & 9 & 11 & 100 & 12 & 17 \\ \hline
si6-r01-m400 & 1 & 66 & 99 & 133 & 89 & 156 & 116 & 50 & 190 & 137 \\ \hline
si6-r01-m800 & 1 & 7 & 235 & 153 & 0 & - & - & 5 & 389 & 125 \\ \hline
si6-r01-m1600 & 1 & 0 & - & - & 0 & - & - & 39 & 499 & 51 \\ \hline
\end{tabular}\\
\end{center}
%
\hspace{-0.7cm}
\begin{tabular}{|l|c|ccccc|cccccc|cccrcc|}
\hline
Bench &  & \multicolumn{ 5}{c|}{CP+Dec} &  \multicolumn{6}{c|}{CP+Dec+h1} & \multicolumn{ 6}{c|}{CP+Dec+h2}   \\ \hline
 & N & \% & $\mu$ & $\sigma$ & D & \#D & \% & $\mu$ & $\sigma$ & D & \#D & S & \% & $\mu$ & $\sigma$ & \multicolumn{1}{c}{D} & \#D & \multicolumn{1}{c|}{S} \\ \hline
si2-r001-m200 & 61E+6 & 94 & 49 & 100 & 91 & 9244 & \textbf{98} & \textbf{6} & \textbf{40} & 98 & 1834 & 0.2 & 87 & 23 & 48 & 71 & 909 & 0.2 \\ \hline
si2-r001-m400 & 17E+8 & 15 & 160 & 177 & 15 & 35655 & \textbf{75} & \textbf{68} & \textbf{125} & 75 & 2268 & 0.4 & 29 & 212 & 218 & 22 & 196 & 0.3 \\ \hline
si2-r001-m800 & 28E+7 & 0 &  -  &  - & 0 & 12 & 4 & 227 & 254 & 4 & 21 & 0.6 & \textbf{12} & \textbf{256} & \textbf{239} & 8 & 0 & 0.6 \\ \hline
si2-r001-m1600 & 2500 & 0 & - & - & 0 & 0 & 7 & 165 & 199 & 1 & 0 & 0.8 & 0 & - & - & 0 & 0 & 0.9 \\ \hline \hline
si6-r01-m200 & 1 & 94 & 148 & 153 & 0 & 0 & \textbf{100} & \textbf{0} & \textbf{0} & 0 & 0 & 1 & 100 & 0 & 0 & 0 & 0 & 1 \\ \hline
si6-r01-m400 & 1 & 2 & 179 & 220 & 0 & 0 & \textbf{100} & \textbf{2} & \textbf{1} & 0 & 0 & 1 & 100 & 4 & 6 & 0 & 0 & 1 \\ \hline
si6-r01-m800 & 1 & 0 &  -  &  - & 0 & 0 & \textbf{100} & \textbf{46} & \textbf{35} & 0 & 0 & 1 & 100 & 46 & 39 & 0 & 0 & 1 \\ \hline
si6-r01-m1600 & 1 & 0 & - & - & 0 & 0 & \textbf{74} & \textbf{479} & \textbf{71} & 0 & 0 & 1 & 54 & 435 & 79 & 0 & 0 & 1 \\ \hline
\end{tabular}
\label{random}
\vspace{-1.5em}
\end{table}

\subsubsection{Analysis -} 

We start the analysis by looking at random graphs (see Table
\ref{random}).  We compare first the \texttt{vflib} with the CP
models \texttt{CPFC} and \texttt{CPAC}.  For \texttt{si2-r001-*}
instances, the \texttt{CPAC} model is the best in mean time and \% of
the solved instances.  When the level of consistency is higher for
the \texttt{MC} constraint, the search space size diminishes, and all
solutions are quickly found.  For \texttt{si6-r01-*} instances,
\texttt{CPAC} is the best model for \texttt{m200} and \texttt{m400}
instances, while \texttt{CPFC} is the best model for \texttt{m800}
and \texttt{m1600} instances.  As shown in Table \ref{degree}, the
mean degree increases with the size of the generated graph. The
effect of propagation is modified.  The \texttt{MC} forward checking
propagator is more efficient with denser graphs than an arc
consistent one.  With sparse graphs, an arc consistent \texttt{MC} is
cheap and propagates a lot, while with denser graphs it is more
efficient to wait for instantiation to propagate.

\begin{table}[tb]
\caption{Irregular meshes, searching for all solutions.}
\begin{center}
\begin{tabular}{|l|r|ccc|ccc|ccc|}
\hline
Bench & \multicolumn{1}{c|}{} & \multicolumn{ 3}{c|}{vflib} & \multicolumn{ 3}{c|}{CPAC} & \multicolumn{ 3}{c|}{CPFC} \\ \hline
 & \multicolumn{1}{c|}{N} & \% & $\mu$ & $\sigma$ & \% & $\mu$ & $\sigma$ & \% & $\mu$ & $\sigma$ \\ \hline
si2-m4Dr6-m625 & 88E+5 & 89 & 23 & 50 & 94 & 21 & 38 & 95 & 6 & 27 \\ \hline
si2-m4Dr6-m1296 & 17E+7 & 16 & 135 & 137 & 33 & 178 & 123 & 38 & 107 & 154 \\ \hline \hline 
si6-m4Dr6-m625 & 3.31 & 100 & 7 & 43 & 100 & 29 & 4 & 100 & 9 & 4 \\ \hline
si6-m4Dr6-m1296 & 10.38 & \textbf{100} & \textbf{13} & \textbf{55} & 100 & 233 & 30 & 100 & 113 & 65 \\ \hline
\end{tabular}
\end{center}
\hspace{-0.3cm}
\begin{tabular}{|l|r|ccccc|cccccc|cccccc|}
\hline
Bench & \multicolumn{1}{c|}{} & \multicolumn{ 5}{c|}{CP+Dec} & \multicolumn{ 6}{c|}{CP+Dec+h1} & \multicolumn{ 6}{c|}{CP+Dec+h2} \\ \hline 
 & \multicolumn{1}{c|}{N} & \% & $\mu$ & $\sigma$ & D & \#D & \% & $\mu$ & $\sigma$ & D & \#D & S & \% & $\mu$ & $\sigma$ & D & \#D & S \\ \hline
si2-m4Dr6-m625 & 88E+5 & 35 & 223 & 151 & 35 & 0.7 & \textbf{100} & \textbf{6} & \textbf{22} & 96 & 5.4 & 0.5 & 94 & 6 & 21 & 88 & 5.5 & 0.3 \\ \hline
si2-m4Dr6-m1296 & 17E+7 & 3 & 120 & 36 & 3 & \multicolumn{1}{r|}{0.1} & \textbf{63} & \textbf{67} & \textbf{109} & 63 & 4 & 0.5 & 49 & 163 & 170 & 49 & 3.9 & 0.5 \\ \hline  \hline
si6-m4Dr6-m625 & 3.3 & 8 & 105 & 32 & 0 & 0 & \textbf{100} & \textbf{7} & \textbf{3} & 6 & 0.1 & 0.8 & 100 & 22 & 26 & 6 & 0.1 & \multicolumn{1}{r|}{0.7} \\ \hline
si6-m4Dr6-m1296 & 10.3 & 0 &  -  &  - & 0 & 0 & 100 & 65 & 20 & 41 & 0.6 & 0.7 & 77 & 223 & 161 & 29 & 0.4 & \multicolumn{1}{r|}{0.7} \\ \hline
\end{tabular}
\label{meshes}
\vspace{-1em}
\end{table}

\begin{table}[tb]
\caption{Mean degree for the tested graph set.}
\begin{center}
\begin{tabular}{|l|cc|}
\hline
Bench & \multicolumn{ 2}{c|}{degree} \\ \hline
 & $\mu$ & $\sigma$ \\ \hline
si2-r001-m200 & 2.30 & 0.14 \\ \hline
si2-r001-m400 & 2.89 & 0.14 \\ \hline
si2-r001-m800 & 3.99 & 0.18 \\ \hline
si2-r001-m1600 & 6.80 & 0.19 \\ \hline \hline
si6-r01-m200 & 3.29 & 0.14 \\ \hline
si6-r01-m400 & 5.27 & 0.16 \\ \hline
si6-r01-m800 & 9.76 & 0.15 \\ \hline
si6-r01-m1600 & 19.20 & 0.17 \\ \hline \hline
si2-m4Dr6-m625 & 3.51 & 0.26 \\ \hline
si2-m4Dr6-m1296 & 3.53 & 0.20 \\ \hline \hline 
si6-m4Dr6-m625 & 5.12 & 0.16 \\ \hline
si6-m4Dr6-m1296 & 5.19 & 0.14 \\ \hline
\end{tabular}
\end{center}
\label{degree}
\vspace{-2em}
\end{table}

We now look at the use of decomposition for random graphs (second
table in Table \ref{random}). 
The first model \texttt{CP+Dec}, which corresponds to a decomposition
approach that uses the whole constraint graph only, fails. This model
cannot take into account the structure of the problem. 
This can be measured through the quality of the decomposition.

First, we will focus on the \texttt{si2-r001-*} classes. 
The models \texttt{CP+Dec+h1} and \texttt{CP+Dec+h2} 
achieve better decompositions than the 
\texttt{CP+Dec} model.
Even though \texttt{CP+Dec} tends to induce more decompositions,
the number of instances using decomposition (see column
D) 
is higher
for \texttt{CP+Dec+h1} and \texttt{CP+Dec+h2} than for
\texttt{CP+Dec}. This visualizes the computational overhead of a pure dynamic
decomposition approach. However, the number of instances using
decomposition tends to be zero for \texttt{m1600} instances.  This is
due to the fact that the graphs have higher degrees as their size
increases (see Table \ref{degree}).  This can be observed by looking
at the column S: the size of the initial subset of variable to
instantiate becomes closer to 100\% as size increases.  For this
reason our decomposition method is beaten by the \texttt{CPAC} model
for \texttt{si2-r001-m1600}.  

We now focus on the \texttt{si6-r01-*}
classes.  As stressed earlier, those instances have denser graphs.
The initial set of variables to instantiate is the whole set of
pattern nodes for \texttt{CP+Dec+h1} and \texttt{CP+Dec+h2}.  No
decomposition occurs.  Why then \texttt{CP+Dec+h*} models outperform
all other methods in those classes?  Because \texttt{CP+Dec+h*}
models use a minsize variable selection policy instead of
\texttt{maxcstr} for \texttt{CPFC}.  
In the class \texttt{si6-r01-*}, the  \texttt{CP+Dec+h1} approach
reduces thus to a \texttt{CPFC} with a minsize variable selection
policy. 

For random graphs, the decomposition method with heuristics is
especially useful for sparse graphs with many
solutions, while a \texttt{CPFC} model using a minsize variable
selection policy seems the best choice for denser graphs and there
are few solutions.  The \texttt{vflib} is clearly outperformed on all
these classes of instances.  Experiments on the other classes of
random graphs, not reported here for lack of space, confirmed this
analysis.  

We now analyze irregular mesh-connected graphs.  We observe in Table
\ref{degree} that the mean degree of the \texttt{si2-m4Dr6-*} classes
is higher than for the \texttt{si6-m4Dr6-*} classes.  We first
compare the \texttt{vflib} and CP models without decomposition.  For
sparser \texttt{si2-m4Dr6-*} classes, \texttt{CPFC} is the best
method, while for denser \texttt{si6-m4Dr6-*} classes, \texttt{vflib}
is the best.  We have no particular explanation for this behavior and
this is an open question.  Regarding decomposition methods, the same
remarks than for random graphs apply.  The \texttt{CP+Dec} model
tends to produce less decomposition than the \texttt{CP+Dec+h*}
models.  Moreover, \texttt{CP+Dec+h*} models are the best models for
sparser instances with many solutions.  As the mean degree of the
instances increase (see Table \ref{degree}), the decomposition
methods become less efficient.  Indeed, for \texttt{si6-m4Dr6-m1296},
the best method is \texttt{vflib}, but our decomposition approach
also solves all the instances and helps CP at diminishing the mean
time.

\subsubsection{Summary -} 

The application of standard direct decomposition methods
\texttt{CP+Dec} lead to performances worse than the direct
application of standard CP models (\texttt{CPFC}, \texttt{CPAC}) and
\texttt{vflib}.  On most classes, the cycle heuristic ({\tt h1}) is
better than the graph partitioning heuristic ({\tt h2}).  On sparse
randomly connected graphs with many solutions, and on sparse
irregular meshes, our decomposition method outperforms standard CP
approaches as well as \texttt{vflib}.  For denser connected graphs,
CP models (\texttt{CPAC} or \texttt{CPFC} with a {\tt minsize}
policy) outperforms \texttt{vflib}.  For denser irregular meshes,
\texttt{vflib}, the standard CP models and our decomposition method
solve all the instances, but \texttt{vflib} is more efficient.

\section{Conclusion}
\label{sec:conc}
Our initial question was to investigate the application of decomposition 
techniques as AND/OR search for problems with global constraints, in particular 
for the SIP. We showed that it is indeed possible using a hybrid approach of 
static and dynamic techniques and a dedicated problem structure analysis. For 
the SIP, one can derive a decomposition enforcing static heuristic that is used 
by a cheap forward checking approach. As soon as the problem gets (likely) 
decomposable, the search process is switched to a fully propagated, dynamically 
decomposed search. This exploits the non-predictable reduction of the 
constraint graph structure via constraint propagation and entailment but reduces 
the huge computational effort of a completely propagated search. We showed that 
our hybrid decomposition approach is able to beat the state-of-the-art 
VF-algorithm for sparse graphs with high solution numbers. As future work, we 
would like to investigate more heuristics for SIP as it influences the quality 
of decomposition. Moreover, we intend to investigate the use of our 
decomposition method for motif discovery where solving SIP is used as an 
enumeration tool~\cite{Grochow:Kellis:motifSearch:RECOMB2007}.


\bibliographystyle{plain}
\bibliography{references}

\end{document}